\documentclass[reprint,superscriptaddress,amsmath,amssymb,prl]{revtex4-1}
\usepackage{graphicx}
\usepackage{dcolumn}
\usepackage{bm}
\usepackage{hyperref}
\hypersetup{colorlinks, citecolor=blue, linkcolor=blue, urlcolor=blue}
\usepackage[dvipsnames]{xcolor}
\usepackage{geometry}
\newcommand{\bb}[1]{\textcolor{Black}{\Large\bfseries #1}}
\newcommand{\newsec}[1]{\paragraph{#1.---}}

\begin{document}
\title{\textnormal{\bb{GALILEO}:
\bb{G}{alactic \bb{A}xion \bb{L}aser \bb{I}nterferometer \bb{L}everaging \bb{E}lectro-\bb{O}ptics} }}
\author{Reza~Ebadi}
\thanks{\href{mailto:ebadi@umd.edu}{ebadi@umd.edu}}
\affiliation{Department of Physics, University of Maryland, College Park, Maryland 20742, USA}
\affiliation{Quantum Technology Center, University of Maryland, College Park, Maryland 20742, USA}
\author{David~E.~Kaplan}
\thanks{\href{mailto:dkaplan@pha.jhu.edu}{dkaplan@pha.jhu.edu }}
\affiliation{The William H. Miller III Department of Physics and Astronomy, The Johns Hopkins University, Baltimore, Maryland 21218, USA}
\author{Surjeet~Rajendran}
\thanks{\href{mailto:srajend4@jhu.edu}{srajend4@jhu.edu}}
\affiliation{The William H. Miller III Department of Physics and Astronomy, The Johns Hopkins University, Baltimore, Maryland 21218, USA}
\author{Ronald~L.~Walsworth}
\thanks{\href{mailto:walsworth@umd.edu}{walsworth@umd.edu}}
\affiliation{Department of Physics, University of Maryland, College Park, Maryland 20742, USA}
\affiliation{Quantum Technology Center, University of Maryland, College Park, Maryland 20742, USA}
\affiliation{Department of Electrical and Computer Engineering, University of Maryland, College Park, Maryland 20742, USA\vspace{4pt}}
\date{\today}

\begin{abstract}
We propose a novel experimental method for probing light dark matter candidates. We show that an electro-optical material's refractive index is modified in the presence of a coherently oscillating dark matter background. A high-precision resonant Michelson interferometer can be used to read out this signal. The proposed detection scheme allows for the exploration of an uncharted parameter space of dark matter candidates over a wide range of masses -- including masses exceeding a few tens of microelectronvolts, which is a challenging parameter space for microwave cavity haloscopes.
\end{abstract}

\maketitle

The nature of dark matter (DM) in modern physics remain elusive. A well-motivated class of DM candidates is light bosonic particles. The QCD axion, for example, is a viable candidate for DM \cite{Preskill:1982cy,Abbott:1982af,Dine:1982ah,Marsh:2015xka,Hook:2018dlk} in addition to solving the Strong CP problem \cite{Peccei:1977hh,Weinberg:1977ma,Wilczek:1977pj}. Axion-like pseudoscalar particles \cite{Marsh:2015xka,Hook:2018dlk} (a generalized form of the QCD axion) and vector particles (e.g., a dark or hidden photon) \cite{Fabbrichesi:2020wbt,Caputo:2021eaa} are similarly well-motivated DM candidates. Such new particles typically have suppressed interactions with the standard model, which nevertheless can be used to search for them in the laboratory \cite{Sikivie:2020zpn,Semertzidis:2021rxs,Caputo:2021eaa,Adams:2022pbo,Rajendran:2022kcs,Sushkov:2023fjw}. 

Light DM is also referred to as wave-like, in contrast to heavier particle-like DM candidates. Due to the high occupancy number of such particles at galactic scales, light DM behaves as a classical wave. Such a DM background can be modeled as a classical random field $a_0\cos(\omega t+\mathbf{k}\cdot\mathbf{x}+\phi)$ \footnote{The oscillatory field will also have a direction if the DM is a vector like dark photon.}, where $a_0=\sqrt{\rho_\mathrm{DM}}/m_\mathrm{DM}$ is the field amplitude given by the DM density $\rho_\mathrm{DM}$ and mass $m_\mathrm{DM}$; $|\mathbf{k}|\simeq m_\mathrm{DM}v$ is the wave number; and $\phi$ is a random phase. The characteristic frequency of the random field's oscillations is given dominantly by the DM mass, with corrections from the kinetic energy, as $\omega\simeq m_\mathrm{DM} + m_\mathrm{DM}v^2/2$, where $v\sim10^{-3}$ is the virial velocity in the Milky Way. The light DM field is therefore coherent over spatial separation $\lambda_\mathrm{c} \sim (m_\mathrm{DM}v)^{-1}$ and over a time scale $\tau_\mathrm{c}\sim(m_\mathrm{DM}v^2)^{-1}$, expressed in natural Planck units \cite{Foster:2020fln}.

Several experimental programs are underway or proposed to probe the parameter space of light DM, with different methods sensitive to specific couplings to standard model physics and a particular range of DM masses. Interactions with the standard model gluons and fermions can be probed via measuring its induced oscillatory electric dipole moments (EDMs) \cite{Abel:2017rtm,Roussy:2020ily,Schulthess:2022pbp}, as well as secondary effects of an oscillatory EDM in precision experiments such as storage rings \cite{Chang:2017ruk,Pretz:2019ham,Kim:2021pld,Alexander:2022rmq,JEDI:2022hxa}, nuclear magnetic resonance \cite{Budker:2013hfa,Aybas:2021nvn}, molecular and atomic spectroscopy \cite{Graham:2011qk,Kim:2022ype}, among others \cite{,Arvanitaki:2021wjk,Berlin:2022mia}. Light DM candidates generically couple to electromagnetism as well, which can be investigated using high-precision methods including resonant cavity haloscopes \cite{ADMX:2021nhd,HAYSTAC:2020kwv,CAPP:2020utb,Jeong:2020cwz,TASEH:2022vvu,CAST:2020rlf,Alesini:2020vny,Quiskamp:2022pks,Cervantes:2022gtv,Tang:2023oid}, lumped elements \cite{Salemi:2021gck,Gramolin:2020ict,Crisosto:2019fcj,DMRadio:2022pkf}, among others \cite{Berlin:2019ahk,Ortiz:2020tgs,Romanenko:2023irv}.

Our lack of knowledge about the nature of DM makes it imperative to probe a wide range of DM parameter space. In addition, different scenarios of cosmological production of the observed DM abundance suggest a wide range of viable masses. For instance, the QCD axion is produced as the pseudo-Nambu-Goldstone boson of spontaneous breaking of the global Peccei-Quinn (PQ) symmetry \cite{Marsh:2015xka,Hook:2018dlk}. Importantly, the QCD axion mass that could serve as DM critically depends on whether the PQ symmetry breaks during inflation or after. Post-inflationary production of the axion serving as DM in principle predicts a unique mass. Even so, it is challenging to solve axion cosmology accurately -- topological defects contribute to axion production on top of the misalignment production, making the dynamics highly nonlinear. Analytical calculations and simulations predict a post-inflationary QCD axion mass that ranges from tens to hundreds of $\mathrm{\mu eV}$ \cite{Ballesteros:2016euj,Berkowitz:2015aua,Bonati:2015vqz,Borsanyi:2016ksw,Dine:2017swf,Klaer:2017ond,Buschmann:2019icd}, with more recent simulations suggesting a narrower range of approximately $40-180\,\mathrm{\mu eV}$ \cite{Buschmann:2021sdq}. A QCD axion with even lower masses are feasible via production pre-inflation and could also serve as DM.

Resonant microwave cavity haloscopes have been the leading DM detectors for $m_\mathrm{DM}\sim\mathrm{\mu eV}$, achieving sensitivity to the QCD axion. However, due to the rapidly diminishing scanning rate caused by a decreasing signal-to-noise ratio (SNR) at smaller cavity volumes \cite{Adams:2022pbo}, it is challenging to probe masses above a few tens of $\mathrm{\mu eV}$ with such detectors. Despite this technical limitation, ongoing efforts are being made to further optimize microwave cavity haloscopes and explore this higher-mass DM parameter space \cite{Caldwell:2016dcw,McAllister:2017lkb,BREAD:2021tpx}, mainly motivated by the post-inflationary axion production as discussed above. 

\begin{figure}
    {\includegraphics[width=0.4\textwidth]{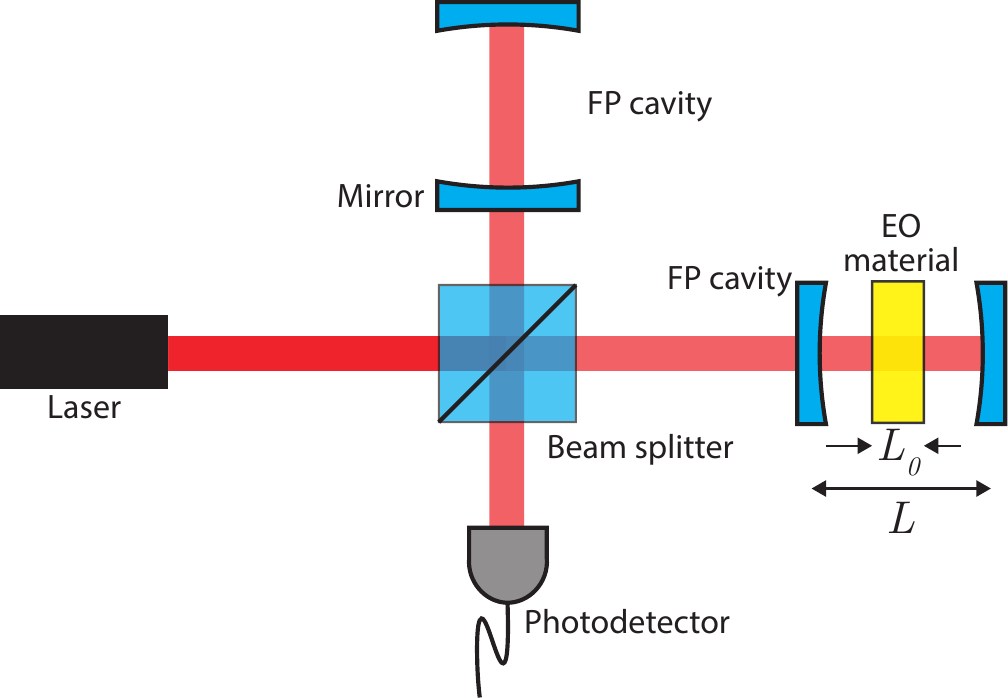}}%
    \caption{Schematic of the proposed laser interferometer-based light dark matter (DM) detector, GALILEO. The Fabry-Perot (FP) cavities are resonant with the light-DM mass $L=2j\pi/m_\mathrm{DM}$. The electro-optical (EO) material's thickness is limited to $L_0\leq\pi/m_\mathrm{DM}$ to preserve the oscillatory DM signal while averaging over laser travel time through the material. Note that the EO material needs to be exposed to a large, uniform magnetic field for axion-induced effects. See text for details.}%
    \label{fig:interferometer}%
\end{figure}

\begin{figure*}
    \centering
    {\includegraphics[width=0.48\textwidth]{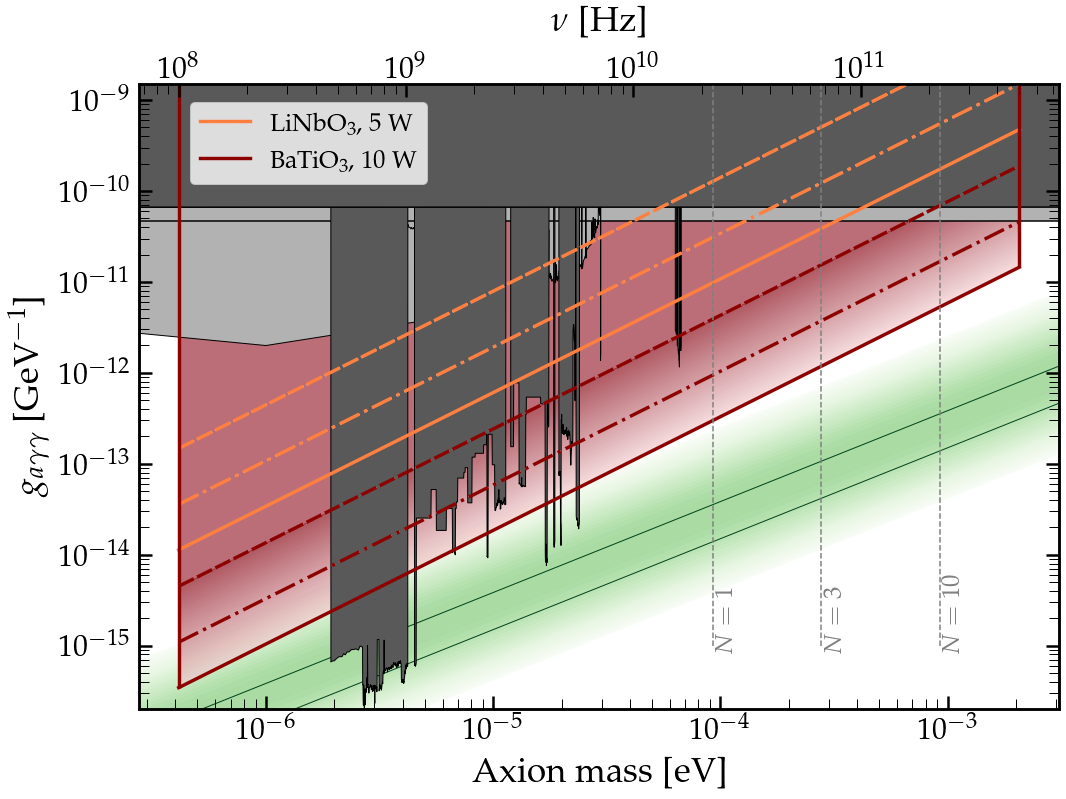}}%
    \qquad
    {\includegraphics[width=0.47\textwidth]{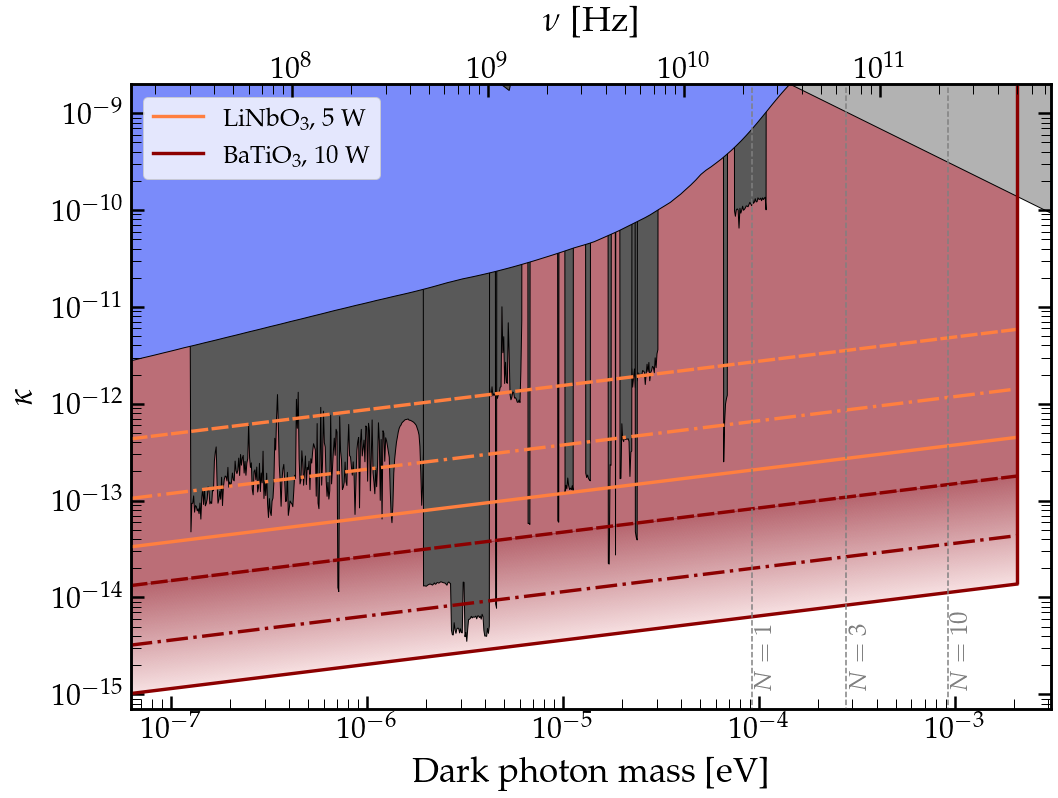}}%
    \caption{Projected sensitivities of the GALILEO experiment for axion (\textit{Left}) and dark photon (\textit{Right}) dark matter searches. The red shaded area is within the reach of the proposed detector. Orange (red) lines: $\mathrm{LiNbO_3}$ ($\mathrm{BaTiO_3}$) as target electro-optical material. Dashed lines: 1\,s averaging at each frequency band $\Delta f=m_\mathrm{DM}/(2\pi\mathcal{F})$. Dash-dotted lines: extended search time of $290$\,s per bin, equivalent to scanning a decade in mass for about 3 years. Solid lines: $290$\,s averaging time per bin and 10 dB squeezing of light input to the interferometer. Vertical gray dashed lines indicate the number of EO material pieces $N=1,3,$ and $10$ needed to achieve maximum sensitivity at representative DM masses if each EO material has a thickness of $L_0=\pi/m_\mathrm{DM}$. See text for details. Dark (light) gray shaded areas are excluded by terrestrial experiments (astrophysical observations). Green: QCD axion parameter space. Blue: excluded by dark photon DM cosmology \cite{Arias:2012az}. Existing limits are adapted from ref.\,\cite{AxionLimits}.}
    \label{fig:sensitivity}
\end{figure*}

In this Letter, we propose a new approach to detect both axion and dark photon DM over a wide mass range, approximately from $0.1-10^3\,\mathrm{\mu eV}$. The basic principle is as follows. Nonlinear electro-optical materials respond to the electric field induced by a coherently oscillating light DM background: the material's refractive index thereby acquires oscillatory corrections. We outline a resonant readout scheme based on laser interferometry to detect such DM-induced signals (Fig.\,\ref{fig:interferometer}); see also Refs.  \cite{Melissinos:2008vn,DeRocco:2018jwe,Obata:2018vvr,Liu:2018icu,Martynov:2019azm,Nagano:2019rbw,PhysRevD.106.115017,Oshima:2023csb,Fedderke:2023dwj} for other interferometry-based DM detection proposals. A Michelson interferometer using a nonlinear electro-optical material in \textit{only} one arm will exhibit an oscillatory differential optical phase between its two arms, imprinted in the measured interferometry fringes. We refer to this experimental approach as GALILEO: Galactic Axion Laser Interferometer Leveraging Electro-Optics. In the following, we present projected sensitivities for this measurement scheme for both the axion and dark photon DM parameter spaces, and compare to the current state of the art (Fig.\,\ref{fig:sensitivity}). Note that for the light DM mass range considered here, the induced oscillations in material refractive index and hence interferometer output signal are in a frequency range $\sim$ 100\,MHz to 1\,THz; and the light-DM background field is coherent over $\mathrm{\lambda_c \sim}$ 1\,cm to 100\,m and $\mathrm{\tau_c \sim 0.1\,\mu s}$ to ms.

\newsec{Electro-optic effect}\label{sec:electrooptic} Light-DM-induced electric fields can be detected via interactions that modulate an electro-optical (EO) material's properties. In particular, the presence of an external electric field results in a change in the polarization of the material, which then modifies the dispersion relation of the electromagnetic wave inside the material. Therefore, one can detect ambient electric fields, such as that induced by light DM, by measuring the effect on propagation of a probe laser through an EO material \cite{Xue:22}.

This scheme requires a nonlinear response of the material that couples the probe laser and the light-DM-induced electric field to be sensed, $E_\mathrm{DM}$. This property can be found in electro-optical materials, where the polarization is given by $P = \epsilon_0\chi^{(1)} E + \epsilon_0\chi^{(2)} E^2+\mathcal{O}(E^3)$ \cite{nonlinearoptics}. Here, $\epsilon_0$ is the vacuum permittivity and $\chi^{(n)}$ is the $n$-th order electric susceptibility of the material \footnote{Note that this equation is written in a schematic form and that it must be computed using proper scalar products.}. We define an effective electric susceptibility as follows: $\chi_\mathrm{eff.} = \chi^{(1)} + \delta\chi$, where $\delta\chi = \chi^{(2)}E_\mathrm{DM}$. Since the effect of DM is expected to be small, one can treat the additional term $\delta\chi$ perturbatively. The electric susceptibility $\chi_\mathrm{eff.}$ is used to calculate a medium's refractive index via $n=(1+\chi_\mathrm{eff.})^{1/2}$. Therefore, in the presence of non-zero $E_\mathrm{DM}$ we have a correction to the material's refractive index proportional to the light-DM-induced electric field $n=\bar{n} + \delta n$, where $\bar{n} = (1+\chi^{(1)})^{1/2}$ and $\delta n \simeq \delta\chi/2\bar{n}$. We calculate this DM-induced refractive index correction for a given set of DM parameters and material properties.

Electro-optic properties due to $\chi^{(2)}$ (i.e., the Pockels effect) can be observed in crystals lacking inversion symmetry. These materials are typically used for applications that employ modulations of the refractive index to achieve fast optical switching and frequency conversion. A widely used example of such a crystal is lithium niobate ($\mathrm{LiNbO_3}$) \cite{LiNbO3_1,LiNbO3_2}, while barium titanate ($\mathrm{BaTiO_3}$) is an emerging material with a higher Pockels coefficient \cite{BTO_nature,BTO_IEEE}. We use these two crystals as benchmarks for the light-DM detector material in our interferometry measurement scheme. The Pockels coefficient $r$ is defined such that the modulation in the refractive index due to an applied electric field is $\delta n = \bar{n}^3 r E/2$. Note that $r$ is a tensor quantity, with its largest component being about $31\,\mathrm{pm/V}$ \cite{LiNbO3_1} ($923\,\mathrm{pm/V}$ \cite{BTO_nature}) for  $\mathrm{LiNbO_3}$ ($\mathrm{BaTiO_3}$). The EO properties of these material are dominantly determined by their lattice structure, in particular the lack of inversion symmetry. Therefore, these values are expected to remain almost the same at low temperatures \cite{youssefi2021cryogenic,LiNbO3atCryo} and/or high magnetic fields as long as there is no significant change to the crystalline structure. Therefore, we have:
\begin{equation}\label{eq:delta_n}
    \delta n \sim 
    \begin{cases}
    1.8\times10^{-10}\,(\mathrm{m/V}) ~E_\mathrm{DM}, &\text{for $\mathrm{LiNbO_3}$}\\
    6.4\times10^{-9}\,(\mathrm{m/V}) ~E_\mathrm{DM}, &\text{for $\mathrm{BaTiO_3}$}
\end{cases}
\end{equation}
where we used $\bar{n}=2.3$ for $\mathrm{LiNbO_3}$ and $\bar{n}=2.4$ for $\mathrm{BaTiO_3}$.

\newsec{Dark matter-induced electric field}\label{sec:Efield} We first consider axion DM coupling to photons:
\begin{equation}
    \mathcal{L} \supset -\dfrac{g_{a\gamma\gamma}}{4}aF_{\mu\nu}\tilde{F}^{\mu\nu} = g_{a\gamma\gamma} a \mathbf{E}\cdot\mathbf{B}
\end{equation}
where $a$ and $F$ are the pseudoscalar axion field and electromagnetic field strength, respectively. This interaction modifies Maxwell's equations. In particular, the axion field generate oscillatory electric and magnetic fields in the presence of a large bias magnetic field $B_0$. When the Compton wavelength of the axion $2\pi/m_\mathrm{a}$ is smaller than the physical size of the magnet, the axion-induced electric field is given by $E_{\rm a} \sim g_{a\gamma\gamma}aB_0$ \cite{Beutter:2018xfx,Ouellet:2018nfr}. Therefore, we have:
\begin{align}\label{eq:Ea}
    E_{\rm a} &\simeq 5.6\times10^{-9}\,{\rm\dfrac{V}{m}} \left(\dfrac{g_{a\gamma\gamma}}{10^{-10}\,{\rm GeV}^{-1}}\right)\left(\dfrac{\rho_\odot}{0.45\,\rm{GeV}/{\rm cm}^3}\right)^{1/2}\nonumber\\
    &\quad\times\left(\dfrac{m_{a}}{100\,{\rm \mu eV}}\right)^{-1}\left(\dfrac{B}{10\,{\rm T}}\right)
\end{align}

For dark photon DM, we consider the kinetic mixing Lagrangian term: 
\begin{equation}
    \mathcal{L}\supset -\dfrac{\kappa}{2}F_{\mu\nu}F'^{\mu\nu}
\end{equation}
where $\kappa$ is the dimensionless mixing parameter and $F'$ is the dark photon field strength. The light-DM-induced electric field due to this mixing term is $E_{\rm d.p.} \sim\kappa\sqrt{\rho_\odot/\epsilon_0}$. Therefore, we have:
\begin{equation}\label{eq:Edp}
    E_{\rm d.p.} \simeq2.8\times10^{-8}\,{\rm\dfrac{V}{m}}\left(\dfrac{\kappa}{10^{-11}}\right)\left(\dfrac{\rho_\odot}{0.45\,\rm{GeV}/{\rm cm}^3}\right)^{1/2}
\end{equation}

\newsec{Detection scheme}\label{sec:detection} We propose using an asymmetric Michelson interferometer, where the sensing volume of the EO material is placed in one arm but not in the other; see Fig.\,\ref{fig:interferometer}. Due to the modulated refractive index, the probe laser will experience a modulated phase velocity as it propagates through the EO material according to $\delta v = -\delta n/\bar{n}^2$. The differential phase velocity between the two arms, integrated over the length of the material, leads to an effective differential arm length $\delta L = \int\mathrm{d}t\,\delta v = -\delta n L_0/\bar{n}$. Hence, the interferometer output will oscillate with the light-DM oscillation frequency. 

Interferometer arms can be equipped with a Fabry-Perot (FP) cavity to further increase the sensitivity via increasing the effective integration length as $L_\mathrm{eff}=L_0 N \mathcal{F}$, where $L_0$, $N$, and $\mathcal{F}$ are the EO material's thickness, number of EO material pieces in the cavity, and finesse of the FP cavity. In order to not average over DM oscillations as the laser beam travels through an EO material of thickness $L_0$within a FP cavity of length $L$, we require that $L_0\leq\pi/m_\mathrm{DM}$ and $L=2j\pi/m_\mathrm{DM}$, in natural Planck units, where $j$ is an integer number. During the experiment, a piezo stage will be used to tune the cavity length, enabling nm-scale spatial movements. We note that in this case, the FP cavities are in resonance with the DM mass and the signal enhancement is due to the laser light traveling repeatedly through the EO material rather than enhancing the DM-induced electric field or the laser amplitude. 

Ultra-high-finesse FP cavities with $\mathcal{F}\sim1.5\times10^5$ are developed for precision experiments \cite{high_finesse_NIST,high_finesse_Pugla,high_finesse_Nicolodi}. Also, FP cavities can achieve Q-factors $\gg10^6$ via extending the cavity length, despite a lower finesse \cite{high_Q_FP,della2014extremely}; assuming $Q\sim10^6$ and $\mathcal{F}\sim10^5$, the cavity length for the parameter space of interest ranges from $\sim\mathrm{cm}$ (higher DM masses) to $\sim\mathrm{m}$ (lower DM masses). Therefore, it is feasible to include multiple EO materials separated by $2\pi/m_\mathrm{DM}$ in a single cavity. The effective travel length through the total amount of EO material is ultimately limited by laser absorption. The absorption coefficient for pure, high-Q nonlinear crystals (including $\mathrm{LiNbO_3}$ \cite{LiNbO3_absorption}) is about $10^{-5}\mathrm{cm^{-1}}$, i.e., $\mathcal{O}(1)$ fraction of the laser power gets absorbed after about $1\,\mathrm{km}$ of traversing inside the EO material. Therefore, we set an upper limit on $L_\mathrm{eff}\leq1\,\mathrm{km}$. Reflection from crystal surfaces is a potential challenge to achieve ultra-high-finesse cavities \cite{ejlli2020pvlas}. However, in our proposed scheme the separations between EO materials and cavity mirrors are set to be multiples of $\pi/m_\mathrm{DM}$, in which case the reflected and transmitted light have the same phase, and therefore the effect of reflection on the cavity finesse can be minimized. Alternatively, high quality anti-reflection (AR) coatings can be used to improve the cavity finesse \cite{zavattini2016polarisation} when there are practical restrictions on the material length such that $L_0\neq\pi /m_\mathrm{DM}$. The use of such an AR coating is especially important in the lower mass regime, where for a high-finesse cavity to maintain $L_0\mathcal{F}\sim1~\mathrm{km}$, the length of the EO material is limited to the sub-optimal case $L_0<\pi/m_\mathrm{DM}$. However, if a state-of-the-art AR coating provides lower finesse, then the length of the material can be chosen to be longer such that $L_0\mathcal{F}$ still remain large. Materials of different lengths can also be used to cover a wider range of parameter space. Optimization of these experimental parameters has yet to be performed, given limitations of the present instrumentation; but it is expected to be straightforward, given the well-established fabrication techniques for EO materials.

Before moving on to computing signal-to-noise (SNR) values for this measurement scheme, we provide the transfer function that relates the light-DM-induced modulation of the refractive index \eqref{eq:delta_n} to the interferometer output signal power:
\begin{equation}\label{eq:transfer_func}
    \dfrac{\delta P_{\rm out}}{\delta n} = \dfrac{\delta P_{\rm out}}{\delta L}\dfrac{\delta L}{\delta n} = \dfrac{2\pi}{\lambda\bar{n}}P_{\rm in} L_0N\mathcal{F}
\end{equation}
where, we used $\delta P_{\rm out}/\delta L = (2\pi/\lambda)P_{\rm in} N\mathcal{F} \sin(8\pi\Delta L/\lambda)$ in the second equality \cite{cahillane2021controlling}. Here, $\lambda$ is the laser wavelength and $\Delta L$ is a DC offset between the two arm lengths, which we choose such that $\delta P_{\rm out}/\delta L$ is maximum, thereby giving optimal sensitivity to a light DM background.

\newsec{Experimental feasibility and projected sensitivities}\label{sec:noise}
Quantum noise and thermal noise are the fundamental sources of noise in the described interferometer measurement scheme. Here, we estimate these two noise sources and show that the proposed experiment can reach the quantum noise limit for experimentally feasible parameters. Technical noise mitigation (such as laser frequency and phase noise, as well as EO material birefringence \cite{della2016pvlas}) is also an important aspect of the final detector, which can benefit from well-established techniques used in state-of-the-art high-precision laser interferometers like LIGO \cite{Abbott:2016xvh,Cahillane:2021jvt,Cahillane:2022pqm}. Note that vibrational noise, a key technical challenge for gravitational wave interferometers, is not a major concern for the proposed DM detector. This is because GALILEO is a high-frequency ($\gtrsim 100\,\mathrm{MHz}$) narrow-band detector, whereas vibrational noise is significant at much lower frequencies. We leave a detailed description of the detector design for a follow-up study. We next discuss the sources of quantum and thermal noise.

Photon counting (shot) noise is the fundamental quantum-mechanical limit of a laser interferometer \cite{Caves_shotnoise}. The number of detected photons follows Poissonian counting statistics, which leads to an output power uncertainty of $\delta P_\mathrm{out} = \hbar\omega_\mathrm{L}\sqrt{N_\mathrm{out}}/\tau$, where $\omega_\mathrm{L}$ is the carrier photon frequency and $N_\mathrm{out}$ is the number of detected photons in the output port over integration time $\tau$. The shot noise amplitude spectral density (ASD) is $\delta P_\mathrm{out}/\sqrt{\Delta f}$, with $\Delta f$ being the bandwidth, which can thus be expressed as: 
\begin{equation}\label{eq:shot_noise}
    \mathrm{ASD_{s.n.}} = \sqrt{2\hbar\omega_\mathrm{L} P_\mathrm{out}} = \sqrt{\hbar\omega_\mathrm{L} P_\mathrm{in}}\,.
\end{equation}

The second fundamental noise source is thermal noise, which has been extensively studied in the context of gravitational wave laser interferometers \cite{Levin:1997kv,Braginsky:1999rp,Braginsky_2000,Evans:2008ei,Dwyer:2014hua}. There are several mechanisms that contribute to the total thermal noise. Homogeneous damping within a material, which is characterized by the imaginary component of Young's modulus, induces interferometer phase noise through elastic deformations of the material. In the presence of inhomogeneous/space-dependent temperature variations, heat flow leads to entropy redistribution and therefore energy dissipation and thermal noise. Such temperature variations can arise from temporal, stochastic fluctuations at a finite temperature or from the photo-thermal effect, i.e., photon absorption inside the material. These fluctuations induce interferometer phase noise via the thermo-elastic effect (due to a non-zero thermal expansion coefficient) and the thermo-refractive effect (due to a non-zero refractive index). In the Supplemental Material, each of the noise sources is estimated using the fluctuation-dissipation theorem (FDT); we find that photon shot noise dominates over thermal noise for temperatures around 200\,K and below. This operational temperature can be achieved even with a few watts of laser absorption (and therefore heat generation) via active cooling feedback \cite{cryo_LIGO}. As shown in refs.\,\cite{youssefi2021cryogenic,LiNbO3atCryo}, the Pockels coefficient is only slightly lower at these operational temperatures than their room temperature values.

In order to project the sensitivity of the GALILEO experiment in the light-DM parameter space, we calculate the shot noise-limited and time-averaged SNR. The DM coherence time $\tau_\mathrm{c}$ plays an important role here. As long as the integration time $t<\tau_\mathrm{c}$, the total number of interferometer signal photons scales linearly with t, the shot noise scales as $\sqrt{t}$, and hence the measurement $\mathrm{SNR\propto \sqrt{t}}$. The SNR degrades for $t>\tau_\mathrm{c}$ as the phase of the light-DM background field varies during the measurement. However, the overall measurement sensitivity to the presence of a non-zero average light-DM background field can still be improved with repeated independent measurements, each lasting for time $\mathrm{\tau_c}$, with the resulting SNR scaling as $\sqrt{T/\tau_\mathrm{c}}$, where $T$ is the total overall time of the repeated measurements. In this repeated measurement regime, the effective noise power spectral density scales as $(T/\tau_\mathrm{c})^{-1/2}$. The SNR scaling behavior in the two regimes can be combined as \cite{Budker:2013hfa}:
\begin{equation}
    {\rm SNR} = \dfrac{\rm \delta P_\mathrm{out}}{\mathrm{ASD_{s.n.}}}(\tau_\mathrm{c} T)^{1/4}
\end{equation}
Using Eqs. \eqref{eq:delta_n}, \eqref{eq:transfer_func}, and \eqref{eq:shot_noise}, we thus estimate GALILEO SNR values for axion and dark photon DM as follows \footnote{We use \eqref{eq:Ea} and \eqref{eq:Edp} as estimates for the DM-induced electric field, neglecting the spatial profile of electric field inside the EO material.}:
\begin{widetext}
\begin{equation}
{\rm SNR} \simeq \left(\dfrac{L_0 N \mathcal{F}}{6.7\,\mathrm{mm}\times1.5\times10^5}\right)\left(\dfrac{\lambda}{1064\,\mathrm{nm}}\right)^{-1/2}\left(\dfrac{P_{\rm in}}{5\,{\rm W}}\right)^{1/2}\left(\dfrac{T}{\mathrm{s}}\right)^{1/4}\times
    \begin{cases}
        20 \left(\dfrac{g_{a\gamma\gamma}}{10^{-10}\,\mathrm{GeV}^{-1}}\right) \left(\dfrac{B}{10\,{\rm T}}\right)\left(\dfrac{m_{\rm DM}}{100\,\mathrm{\mu eV}}\right)^{-5/4}\\
        120 \left(\dfrac{\kappa}{10^{-11}}\right) \left(\dfrac{m_{\rm DM}}{100\,\mathrm{\mu eV}}\right)^{-1/4}
    \end{cases}
\end{equation}
\end{widetext}
where we use $\mathrm{BaTiO_3}$ as the EO material. We set projections in the axion and dark photon DM parameter space using the criterion $\mathrm{SNR\sim1}$, as shown in Fig.\,\ref{fig:sensitivity}. The scanning rate will also be affected by the duty cycle determined by the experimental details, such as speed of mirror movement using nanopositioning piezo stages and cavity stabilization. We leave a detailed examination of experimental parameters to a future study, presenting the best theoretical projections here.

To achieve maximum sensitivity in the higher-mass regime, we propose using multiple EO materials inside a FP cavity, each separated by $2\pi/m_\mathrm{DM}$. As discussed above, laser absorption in the EO material limits $L_\mathrm{eff}\leq\mathrm{1\,km}$. This means that for $\mathcal{F}=1.5\times10^5$ we have $L_0N\lesssim\mathrm{6.7\,mm}$. Therefore, for $m_\mathrm{DM}\lesssim\mathrm{90\,\mu eV}$ (corresponding to $L_0\gtrsim\mathrm{6.7\,mm}$) we use a single EO material with a thickness of $\mathrm{6.7\,mm}$; whereas for $m_\mathrm{DM}\gtrsim\mathrm{90\,\mu eV}$ (corresponding to $L_0=\pi/m_\mathrm{DM}<\mathrm{6.7\,mm}$) we use multiple EO materials. Vertical gray dashed lines in Fig.\,\ref{fig:sensitivity} indicate the required number of EO materials for the higher-mass DM parameter space to achieve maximum sensitivity.

In the proposed detection scheme, the background DM-induced electric field oscillations manifest as an oscillatory signal at the interferometer output. It is therefore crucial to have a high photodetector bandwidth in order to resolve higher-mass DM-induced oscillations. While commercially available low-noise photodetectors have a bandwidth of up to 50 GHz (corresponding to $m_\mathrm{DM}\simeq\mathrm{210\,\mu eV}$), there are currently demonstrations of detectors with bandwidths up to 500 GHz \cite{ultrafast_PD,ultrafast_PD_2}. The EO material's response time will also limit detection of DM masses higher than a few $\mathrm{meV}$. We thus set a higher mass limit of about 2 meV in Fig.\,\ref{fig:sensitivity}, corresponding to a photodetector bandwidth of about 500\,GHz. In the low-mass axion regime the Fabry-Perot cavity length becomes a limiting factor, because at least one arm of the interferometer must be within the magnet producing the large bias magnetic field necessary for axion-induced signals. Therefore, we consider only axion masses greater than $\mathrm{0.4\,\mu eV}$ in the sensitivity estimations. This requirement is more relaxed for dark photon searches, where no background magnetic field is needed.

It is possible to reduce the observed noise below the nominal shot-noise limit  through \textit{squeezing}, where the electromagnetic vacuum noise in the measurement readout quadrature is reduced, with a corresponding increase of the noise in the other quadrature, consistent with the Heisenberg quantum limit. To date, laser interferometric gravitational wave detectors have successfully achieved 10\,dB vacuum squeezing \cite{10dB_squeezing,11dB_squeezing,3dB_squeezing}, which is equivalent to improved sensitivity by a factor of about 3. As part of our sensitivity projections in Fig.\,\ref{fig:sensitivity}, we also consider squeezing to further improve the detector reach, assuming similar performance (including losses) as achieved in interferometer-based gravitational wave detectors.

\newsec{Summary}\label{sec:discussion} We proposed a new approach to detect axion and dark photon dark matter (DM) over almost four decades in mass from about $0.1-10^3\,\mathrm{\mu eV}$. We dub this experiment GALILEO, which is based on laser interferometry and uses electro-optical properties to detect DM-induced electric fields. The proposed experiment explores parameter spaces that are challenging to probe with resonant cavity haloscopes in the high mass and low mass regimes. In the higher mass regime of the axion parameter space, our projected sensitivity is similar to the current limits from resonant cavity experiments \cite{Quiskamp:2022pks} at about 16 GHz with the same integration time, with promising prospects to explore even higher-mass candidates. In the lower mass regime of the axion parameter space, the geometric requirements are simpler to satisfy for GALILEO, since only one arm of the interferometer with length $\propto 1/m_\mathrm{a}$ needs to be inside a high field magnet. Future technical improvements, such as the development of materials with enhanced electro-optical properties, may extend the reach of this approach to the QCD axion dark matter parameter space across a range of several orders of magnitude for the axion mass. A dark photon DM search with GALILEO is also promising, since it enables probing unexplored parts of parameter space with more relaxed design requirements, e.g., with no need for a high field magnet or with lower Q-factor FP cavities.

\vspace{10pt}
\textit{We are grateful to John~W.~Blanchard, Johannes~Cremer, Anson~Hook, Jner~Tzern~(JJ)~Oon, Aakash~Ravi, and Erwin~H.~Tanin for insightful discussions.  We also thank Alexander Millar and Giuseppe Ruoso for helpful comments on the preprint. This work was supported by the Argonne National Laboratory under Award No. 2F60042; the Army Research Laboratory MAQP program under Contract No. W911NF-19–2-0181; the DOE fusion program under Award No. DE-SC0021654; and the University of Maryland Quantum Technology Center. This work was also supported by the U.S.~Department of Energy (DOE), Office of Science, National Quantum Information Science Research Centers, Superconducting Quantum Materials and Systems Center (SQMS) under contract No.~DE-AC02-07CH11359. D.E.K. and S.R. are supported in part by the U.S. National Science Foundation (NSF) under Grant No. PHY-1818899. S.R. is also supported by the DOE under a QuantISED grant for MAGIS, and the Simons Investigator Award No. 827042. This work was also made possible through the support of Grant 63034 from the John
Templeton Foundation. The opinions expressed in this publication are those of the authors and
do not necessarily reflect the views of the John Templeton Foundation.}

\bibliography{references}

\end{document}